\documentclass[10pt,conference,final,letterpaper]{IEEEtran}
\pdfoutput=1
\IEEEoverridecommandlockouts
\usepackage{cite}
\usepackage{nicefrac}
\usepackage{amsmath,amssymb,amsfonts}
\usepackage{algorithmic}
\usepackage{graphicx}
\usepackage{textcomp}
\usepackage{xcolor}
\usepackage{orcidlink}
\usepackage{tikz}
\usetikzlibrary{decorations.pathreplacing,calligraphy}

\usepackage{subcaption}
\def\BibTeX{{\rm B\kern-.05em{\sc i\kern-.025em b}\kern-.08em
    T\kern-.1667em\lower.7ex\hbox{E}\kern-.125emX}}
    
\begin{document}
\bstctlcite{IEEEexample:BSTcontrol}

\title{Towards Transfer Learning for Large-Scale Image Classification Using Annealing-based Quantum~Boltzmann~Machines\\
\thanks{© 2023 IEEE.  Personal use of this material is permitted.  Permission from IEEE must be obtained for all other uses, in any current or future media, including reprinting/republishing this material for advertising or promotional purposes, creating new collective works, for resale or redistribution to servers or lists, or reuse of any copyrighted component of this work in other works.\\ The authors acknowledge funding from the German Federal Ministry for Economic Affairs and Climate Action, project PlanQK, 01MK20005I.}
}

\makeatletter
\newcommand{\linebreakand}{
  \end{@IEEEauthorhalign}
  \hfill\mbox{}\par
  \mbox{}\hfill\begin{@IEEEauthorhalign}
}
\makeatother

\author{
\IEEEauthorblockN{Daniëlle Schuman\textsuperscript{$\orcidlink{0009-0000-0069-5517}$}}
\IEEEauthorblockA{\textit{LMU Munich}\\
danielle.schuman@ifi.lmu.de}
\and
\IEEEauthorblockN{Leo Sünkel}
\IEEEauthorblockA{\textit{LMU Munich}\\
leo.suenkel@ifi.lmu.de}
\and
\IEEEauthorblockN{Philipp Altmann\textsuperscript{$\orcidlink{0000-0003-1134-176X}$}}
\IEEEauthorblockA{\textit{LMU Munich}\\
philipp.altmann@ifi.lmu.de}
\and
\IEEEauthorblockN{Jonas Stein\textsuperscript{$\orcidlink{0000-0001-5727-9151}$}}
\IEEEauthorblockA{\textit{LMU Munich}\\
jonas.stein@ifi.lmu.de}
\linebreakand
\IEEEauthorblockN{Christoph Roch\textsuperscript{$\orcidlink{0000-0003-0781-6590}$}}
\IEEEauthorblockA{\textit{LMU Munich}\\
christoph.roch@ifi.lmu.de}
\and
\IEEEauthorblockN{Thomas Gabor}
\IEEEauthorblockA{\textit{LMU Munich}\\
thomas.gabor@ifi.lmu.de}
\and
\IEEEauthorblockN{Claudia Linnhoff-Popien\textsuperscript{$\orcidlink{0000-0001-6284-9286}$}}
\IEEEauthorblockA{\textit{LMU Munich}\\
linnhoff@ifi.lmu.de}
}

\maketitle

\begin{abstract}
Quantum Transfer Learning (QTL) recently gained popularity as a hybrid quantum-classical approach for image classification tasks by efficiently combining the feature extraction capabilities of large Convolutional Neural Networks with the potential benefits of Quantum Machine Learning (QML). Existing approaches, however, only utilize gate-based Variational Quantum Circuits for the quantum part of these procedures. In this work we present an approach to employ Quantum Annealing (QA) in QTL-based image classification. Specifically, we propose using annealing-based Quantum Boltzmann Machines as part of a hybrid quantum-classical pipeline to learn the classification of real-world, large-scale data such as medical images through supervised training. We demonstrate our approach by applying it to the three-class COVID-CT-MD dataset, a collection of lung Computed Tomography (CT) scan slices. Using Simulated Annealing as a stand-in for actual QA, we compare our method to classical transfer learning, using a neural network of the same order of magnitude, to display its improved classification performance. We find that our approach consistently outperforms its classical baseline in terms of test accuracy and AUC-ROC-Score and needs less training epochs to do this.
\end{abstract}

\begin{IEEEkeywords}
quantum transfer learning, quantum annealing, simulated annealing, quantum machine learning, quantum boltzmann machine
\end{IEEEkeywords}

\section{Introduction}

Promising great advantages such as speed-ups, increased space efficiency and the ability to model probability distributions naturally, quantum computing is being heavily investigated for usage in machine learning (ML) applications~\cite{adcock2015advances, o2020quantum, schuld2018supervised}. To combine these potential advantages with the ability to efficiently extract “highly informative features”~\cite[p.~1]{mari2020transfer} from large-scale image data using pre-trained, state-of-the-art classical Convolutional Neural Networks (CNNs), hybrid Quantum Transfer Learning (QTL) has recently seen a surge in popularity in such applications~\cite{mari2020transfer, sequent, pramanik2022quantum, otgonbaatar2022quantum, azevedo2022quantum, majumdar2023histopathological, umer2022integrated, acar2021covid, leo}. These approaches combine said classical CNNs with Variational Quantum Circuits (VQCs) trained on the data to be learned in a supervised manner~\cite{mari2020transfer, sequent, pramanik2022quantum, otgonbaatar2022quantum, azevedo2022quantum, majumdar2023histopathological, umer2022integrated, acar2021covid, leo}. They often achieve desirable results, even though they use only the small qubit count available on the current Noisy Intermediate-Scale Quantum (NISQ) computers of the gate-model type~\cite{sequent, mari2020transfer, umer2022integrated, acar2021covid, Preskill2018quantumcomputingin}. Another type of still noisy~\cite{mcgeoch2020theory}, but currently available Quantum Computers (QCs), namely Quantum Annealers, can also be used for supervised QML: This happens e.g. in the form of Quantum Boltzmann Machines (QBM)~\cite{Amin18}, which might offer a speed-up and potential improvement of the training process~\cite{Adachi15, kurowski2021applying, dixit2020training, Amin18}. 
Eager to explore the applicability of Transfer Learning~(TL) in Quantum Annealing~(QA), we want to evaluate the compatibility of QBMs with QTL approaches, to see if we can combine the advantages of both concepts.
 
To do this, we propose an approach that combines the QTL-method SEQUENT~\cite{sequent}, a two-step process that connects a CNN to a quantum classifier using a classical compression layer, with an annealing-based QBM that takes the role of said classifier. We test this approach by classifying the COVID-CT-MD dataset~\cite{COVID-CT-MD}, a large-scale, real-world, three-class medical image dataset consisting of slices of lung Computed Tomography (CT) scans. To save QPU time, we do this using the Simulated Annealing algorithm (SA)~\cite{Siddique16} instead of QA, and compare this approach to a classical TL one using a similarly-sized Feed-forward Neural Network (FNN).

The rest of the paper is structured as follows: In Sec.~\ref{sec:related}, we introduce QBMs as well as QTL, and mention the most relevant work in the area. We then present our architecture in Sec.~\ref{sec:method}, and describe the COVID-CT-MD dataset~\cite{COVID-CT-MD} and our experiments performed on it in Sec.~\ref{sec:experiments}. We conclude by discussing the implications and limitations of our work, as well as the needs for future research derived from it, in Sec.~\ref{sec:conclusion}.

\section{Background and Related Work}\label{sec:related}
\subsection{(Quantum) Boltzmann Machines}
A \emph{Boltzmann Machine (BM)} is an undirected, arbitrarily connected, stochastic neural network (NN) whose neurons $s_i \in \mathbf{s}$, called visible and hidden units, can take the values 0 and 1 with a certain probability~\cite{ackley1985learning}. The function modeled by the BM takes the form of a Boltzmann distribution governing said probabilities~\cite{ackley1985learning, Benedetti16, stein2023exploring}:
\begin{equation}
    P_{\text{model}}(\mathbf{s}) = \dfrac{e^{E(\mathbf{s})}}{\sum_\mathbf{s} e^{E(\mathbf{s})}} \text{ with } E(\mathbf{s}) = \sum_{ij} w_{ij} s_i s_j + \sum_i b_i s_i
\end{equation}
where $w_{ij}$ and $b_i$ are the networks weights and biases and $E$ is the \emph{energy function}.
The BM can be trained using stochastic gradient descent to minimize the Kullback-Leibler divergence ($D_{KL}$) between $P_{\text{model}}$ and the distribution $P_{\text{data}}$ underlying the dataset to be learned, for which, in case of supervised learning, the conditional distribution of the datapoints' labels given their input vectors is used~\cite{ackley1985learning, Amin18}. Training involves sampling from both distributions, as the gradient of $D_{KL}$ is~\cite{ackley1985learning, stein2023exploring}:
\begin{align} \label{eq:kl-gradient}
    \partial_{w_{ij}} D_{KL}\left(P_{\text{data}}\mid\mid P_{\text{model}}\right) =& \langle s_i s_j \rangle_{\text{data}} - \langle s_i s_j \rangle_{\text{model}} \\
    \partial_{b_{i}} D_{KL}\left(P_{\text{data}}\mid\mid P_{\text{model}}\right) =& \langle s_i  \rangle_{\text{data}} - \langle s_i \rangle_{\text{model}}
\end{align}

where $\langle . \rangle$ denotes averaging samples. Sampling thus means determining the values of all units $s_i$ and $s_j$ that are not \emph{clamped}, i.e. fixed to values of a datapoint, several times, once for each sample. In supervised learning, one always clamps some of the visible units to the values of an input vector, and sampling from $P_{\text{data}}$ additionally involves clamping the rest of them to the label~\cite{ackley1985learning, Amin18}. The difference between classical BMs and QBMs lies in the way the values of the unclamped units are obtained: Classically, one has to calculate the value of each unit based on its probability of becoming 1 of
\begin{equation}\label{eq:prob}
    p (s_i = 1) = \left(1 + e^{\sum_j w_{ij} s_j + b_i}\right)^{-1}
\end{equation}
which depends on the current values of its neighbors $s_j$~\cite{ackley1985learning, fischer2012introduction}. This is done iteratively for all units in a sample until their values do not change anymore~\cite{ackley1985learning, fischer2012introduction}. As this has to be done for every sample taken for every datapoint in every epoch, the process can become very time consuming~\cite{ackley1985learning, fischer2012introduction}.

This is why general classical BMs are only rarely used nowadays\footnote{ This can be e.g. seen when searching Google Scholar for recent literature on BMs while excluding results containing “restricted” and “quantum”: Most of the few fitting results consider either niche scientific applications or special types of BMs, such as Deep BMs, chaotic BMs or higher-order BMs.}. Instead, one mainly uses \emph{Restricted BMs (RBMs)} which make calculations more efficient by  restricting their connectivity to enable parallel calculations of some units' values and approximating samples from $P_{\text{model}}$ by calculating them using samples from $P_{\text{data}}$ ~\cite{upadhya2019overview, montufar2018restricted, hinton2012practical}. 
QBMs on the other hand can sample the values of all unclamped units at once, without any restrictions or approximations~\cite{Amin18}. This is possible as current Quantum Annealers do in general return samples for entire bit-vectors $\mathbf{s}$ that are approximately Boltzmann distributed with regards to an energy function that they are given in the form of e.g. a Quadratic Unconstrained Binary Optimization (QUBO) problem~\cite{Amin18}. Thus, mapping each unclamped unit to a logical qubit of the annealer, one can directly map a BMs weights and biases to a QUBO, provided they are scaled with the appropriate inverse effective temperature $\beta_{\textrm{eff}}$ of the hardware, and determine the units values using one run of the QA process per sample~\cite{Amin18, Benedetti16}.

Thus, previous work like \cite{dixit2020training, kurowski2021applying, Adachi15} and \cite{dorband2016boltzmann} that used standalone QA-based QBMs trained with supervised learning for image classification has observed benefits in terms of computation time~\cite{Adachi15, kurowski2021applying} or less fluctuations in training accuracy~\cite{dixit2020training} in comparison to classical equivalents. However, these papers usually classified only rather simple datasets like the “bars-and-stripes”~\cite{dixit2020training} or the MNIST dataset~\cite{Adachi15, kurowski2021applying, dorband2016boltzmann}, where extensive feature extraction as provided by a TL approach using a large CNN is not necessary. The only annealing-based work similar to ours is \cite{piat2018image}, which uses a pipeline of a classical autoencoder and a Deep Belief Network pre-trained with QA-based restricted QBMs to classify, among others, medical images. Unlike this approach, we train our QBM to directly classify the images using supervised learning, instead of only using it to initialize a classical classifier.

\subsection{(Quantum) Transfer Learning}

\emph{Transfer learning (TL)} is a ML technique which re-uses previously learned knowledge by adapting an existing ML model $M_A$ (usually a large NN) which has been trained on a dataset $D_A$ to perform a similar learning task on a usually similar dataset $D_B$~\cite{Morid_2021, sequent, leo}. 
This is done by replacing part of $M_A$, in case of a large CNN usually the last (few) fully-connected layers used for classification, by a new comparable component such as an untrained classifier $M_B$ (often also consisting of fully connected layers)~\cite{godasu2020transfer, Morid_2021, leo}. The resulting new model $M_{A'B}$ is subsequently trained on the new dataset $D_B$ while either freezing the (usually convolutional) layers $M_{A'}$ taken from $M_A$, meaning their weights and biases are not trained, or just training them along with $M_B$~\cite{mari2020transfer, sequent, leo, Morid_2021, godasu2020transfer}. The first approach, which is used in this work, is called \emph{feature extraction}, as that is what $M_A'$ is used for, the second is called \emph{fine-tuning}~\cite{godasu2020transfer, Morid_2021, sequent}. Both have the advantage that less datapoints are needed in $D_B$ to achieve a model with good generalization and less computational resources are needed compared to training a model $M_{A'B}$ from scratch~\cite{kim2022transfer, godasu2020transfer, Morid_2021, sequent}. 

In Quantum Transfer Learning (QTL), the same methods are applied to hybrid QML models, meaning that either $M_{A'}$ or $M_B$ or both, contain or consist entirely of a QML component such as a trainable VQC~\cite{mari2020transfer}. Using a classical $M_A$ and a hybrid quantum-classical $M_B$ is particularly popular, as employing a classical $M_{A'}$ as a feature extractor allows to drastically downsample inputs, making even large datapoints small enough to be processed on a NISQ computer~\cite{leo}.

VQC-based QTL approaches that process large datasets, like the COVID-CT-Scans, on gate-model-type QCs are \cite{pramanik2022quantum}, \cite{otgonbaatar2022quantum}, \cite{azevedo2022quantum}, \cite{majumdar2023histopathological}, \cite{umer2022integrated}, \cite{acar2021covid} and \cite{leo}. 
A gate-based QTL approach especially relevant to our work is SEQUENT~\cite{sequent}, a two-step TL approach. Using a large pre-trained CNN like ResNet-18, it first replaces the fully-connected layer(s) at the end of this network with a classical \emph{compression layer}, for further down-sampling data, and a surrogate classical classification layer~\cite{sequent}. These layers are both trained on a dataset $D_B$ while freezing the other layers of the CNN~\cite{sequent}. Then, in a second step, the surrogate classification layer is replaced by a VQC, which is trained while the rest of the network, including the compression layer, is being frozen~\cite{sequent}. This way of training allows the impact of this quantum part on the classification performance to become more visible~\cite{sequent}.\\

\section{Methodology} \label{sec:method}

\begin{figure*}[t!]
\centerline{\includegraphics[width=\textwidth]{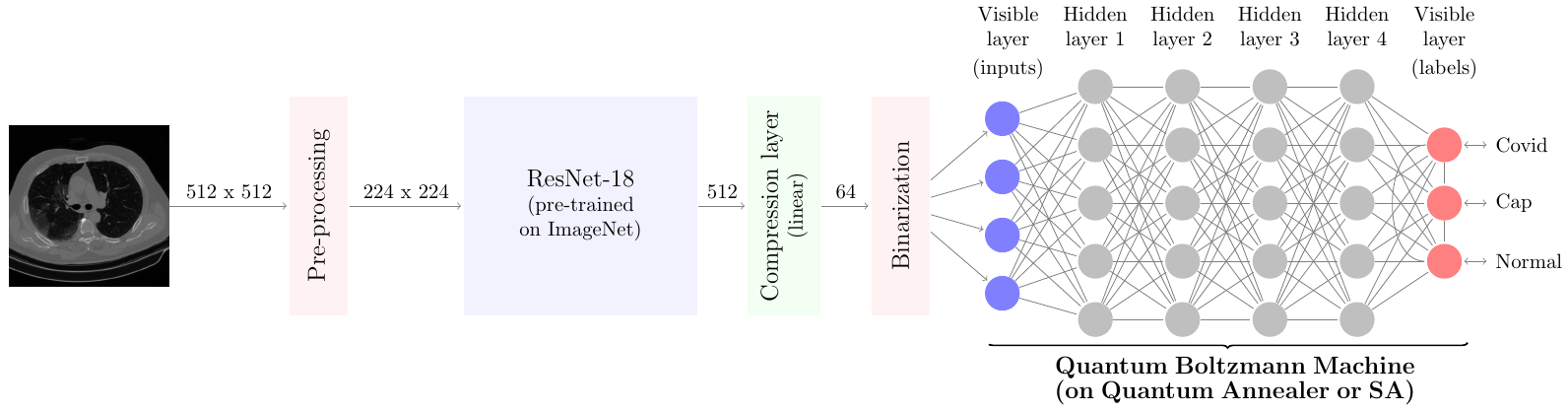}}
\caption{Our hybrid QTL pipeline. The QBM is the only part of this pipeline that can be run on a Quantum Annealer (even though we used SA in our experiments), the rest is executed on a classical computer. (CT image from \protect \cite{COVID-CT-MD})}
\label{fig:pipeline}
\end{figure*}

Our approach for constructing a QA-based QTL pipeline for large-scale image classification, which can be seen in Fig.~\ref{fig:pipeline}, follows a process very similar to SEQUENT~\cite{sequent}, only replacing the VQC with an annealing-based QBM. 

It starts with a pre-processing phase in which we resize and subsequently crop the input images using functionalities of the \textit{Pytorch} library~\cite{pytorch} to scale them from their original size, e.g. 512 x 512 pixels for our COVID-CT-MD dataset~\cite{COVID-CT-MD}, to a size of 224 x 224 pixels.
After pre-processing, the image is fed into a ResNet-18~\cite{resnet} which was pre-trained on the \textit{ImageNet} data set~\cite{imagenet} for feature extraction. This network's last layer has been replaced by a compression layer which further down-samples the data from dimensionality 512 to 64, and has been pre-trained for 10 epochs on our COVID-CT-MD dataset~\cite{COVID-CT-MD} using a surrogate classification layer. This layer has subsequently been replaced by a layer that binarizes the data and a (deep) QBM, which classifies the data point into one of our three categories. In our experiments, the number of hidden layers $h$ and the total number of hidden units $n$ of this (deep) QBM were treated as hyperparameters to be optimized, $h$ ranging from 1 to 4 and $n$ ranging from 12 to 500. These ranges were chosen to allow the exploration of many combinations of $h$ with different numbers of units per layer, while ensuring that most of these combinations could potentially be embedded into current QA hardware~\cite{quantum_hardware}.

Regarding the rationale behind using this two-step TL approach, we would like to point out that, being an undirected neural network, a QBM cannot be trained using backpropagation. This makes it impossible to train any upstream classical feed-forward layers of a combined neural network architecture simultaneously. Thus, using a SEQUENT-like approach to train any compression layers is in this case not only advantageous regarding the investigation of the impact of this quantum-part on the network, it is outright unavoidable if one wants to employ such a layer.\\

\section{Experiments} \label{sec:experiments}

To show that this architecture can be used for large-scale image classification, we perform experiments on the COVID-CT-MD dataset~\cite{COVID-CT-MD}. Like Stein et al.~\cite{stein2023exploring}, we do this using SA instead of actual QA, as the hyperparameter optimization we perform in our experiments requires an extremely large amount (hundreds of millions\footnote{One arrives at this order of magnitude by multiplying 55 runs * 10 seeds * 2715 training data points * 2 sampling phases (from $P_{\text{data}}$ and $P_{\text{model}}$) * between 1 and 20 epochs * between 5 and 100 samples per sampling phase.}) of annealing runs, which is currently not feasible for us to execute on quantum hardware given the scarce availability of QA machines. We consider it to be reasonable to use SA as a stand-in for QA in this context, even though it has slightly different working mechanisms due to the absence of quantum tunneling in the algorithm and might take a lot more execution time on certain problem instances~\cite{crosson2016simulated, denchev2016computational, nishimori2015comparative}, as it also returns approximately Boltzmann-distributed results, just like QA~\cite{Siddique16, eglese1990simulated, Amin18}. This is also stated in the documentation of D-Wave's SA implementation~\cite{D-WaveSAS}, which we use in our experiments.

\subsection{Dataset}

The COVID-CT-MD dataset~\cite{COVID-CT-MD} consists of lung CT scan slices labeled as COVID-19 pneumonia (“Covid”), Community Acquired Pneumonia (“Cap”) and healthy (“Normal”)\footnote{In contrast to both other classes, the “Normal” images are not labeled per slice, but only per patient. Hence, we here selected random slices between the indexes 15 and 112, in most of which the lung takes up a decent part of the image.} which we convert from DICOM format to gray-scale PNG images sized 512 x 512 pixels. We split this dataset into a train and a test set which each contain a small, but equal amount of patients for each class (20 for the train set, 5 for the test set), the number of which is limited by the available amount of Cap patients (which is 25). We also balance the sets regarding the total amount of images per class, by deleting images as necessary, always selecting the patients of whom we had the most slices when doing so. This results in 905 images per class in the training and 275 images per class in the test set. In order to be able to increase the size of the training and test sets regarding the number of patients, as well as to save valuable computation time running the SA or, in the future, QA algorithm, we refrain from using validation.

\subsection{Experimental setup and results}

The first step of our experiments is to perform a hyperparameter optimization, using the Bayesian search algorithm of the \textit{weights and biases framework}~\cite{wandb} to find hyperparameters that maximize the average of training accuracy and training AUC-ROC-Score. The hyperparameters being optimized can be taken from the top line of Table~\ref{tab1}. We continue this process for 55 runs, each time averaging over 10 random seeds.

To assess the benefits of using an annealing-based QBM in this pipeline, we compare our approach to a classical one using the same pipeline, but replacing the QBM with a simple FNN using sigmoidal units\footnote{The sigmoid activation function was chosen due to being almost equivalent to (\ref{eq:prob})~\cite{fischer2012introduction}, causing it to give the neurons comparable outputs to the average values of units in a BM and thus maximizing the similarity of both approaches.}, with a similar amount of parameters to optimize (again having the number of its hidden units be subject to optimization). 

We chose this as a classical baseline instead of a classical (R)BM since it falls into the category of backpropagation-based approaches, which are much more commonly used in state-of-the-art medical image classification techniques~\cite{kim2022transfer, sequent}. Furthermore, FNNs are one of the most common options for this type of component in classical feature-extraction-based transfer learning for medical image classification~\cite{godasu2020transfer, Morid_2021}, making them suitable our setting. This baseline was also subjected to the same hyperparameter optimization procedure.

\begin{table*}[htbp]
\caption{Best identified hyperparameter settings (rounded to 5 decimal places). $\beta_{\textrm{eff}}$ and the sample count are not applicable to the learning procedure of the classical approach. $\mathbf{\beta_1}$, $\mathbf{\beta_2}$ and $\mathbf{\epsilon}$ are parameters of the Adam optimizer\protect~\cite{kingma2014adam} we used.}
\centering
\begin{tabular}{|c|c|c|c|c|c|c|c|c|c|c|c|}
\hline
\textbf{approach}& \textbf{name}& \textbf{batch size}& \textbf{epochs}& \textbf{$\mathbf{h}$}& \textbf{$\mathbf{n}$}& \textbf{learning rate }& \textbf{Adam's $\mathbf{\beta_1}$}& \textbf{Adam's $\mathbf{\beta_2}$}& \textbf{Adam's $\mathbf{\epsilon}$}& \textbf{$\mathbf{\beta_{\text{eff}}}$}& \textbf{sample count}\\
\hline
SA & ba\_86 & 86 & 11 & 1 & 332 & 0.02810 & 0.57372 & 0.87481 & 0.46784 & 5.57229 & 60 \\
\hline
SA & ba\_87 & 87 & 8 & 1 & 253 & 0.05893 & 0.57614 & 0.85976 & 0.57921 & 7.33457 & 51\\
\hline
SA & ba\_99 & 99 & 6 & 1 & 492 & 0.06456 & 0.73595 & 0.89025 & 0.61694 & 9.31536 & 94\\
\hline
classical & ba\_29 & 29 & 13 & 1 & 279 & 0.05120 & 0.53482 & 0.87913 & 0.22486 & - & -\\
\hline
classical & ba\_35 & 35 & 15 & 1 & 294 & 0.02592 & 0.57760 & 0.80141 & 0.16210 & - & -\\
\hline
classical & ba\_77 & 77 & 18 & 1 & 82 & 0.09008 & 0.71723 & 0.83271 & 0.10328 & - & -\\
\hline
\end{tabular}
\label{tab1}
\end{table*}

When plotting the training accuracy values and AUC-ROC-Scores of the resulting three best identified hyperparameter settings for both of these models, listed in Table~\ref{tab1} and shown in Fig.~\ref{fig:train}, one can see that the annealing-based approach not only reaches significantly higher values, training results are also significantly more consistent over different training seeds compared to the classical approach. And while the training with SA does take far more wall clock time compared to the classical baseline, the figure also shows it executes less epochs in said time to reach this performance. This might lead to a future speed-up in training if using actual QA were to greatly reduce the time needed per training step in comparison to SA.

\begin{figure*}[htbp]
  \begin{subfigure}{\columnwidth}
  \includegraphics[width=\textwidth]{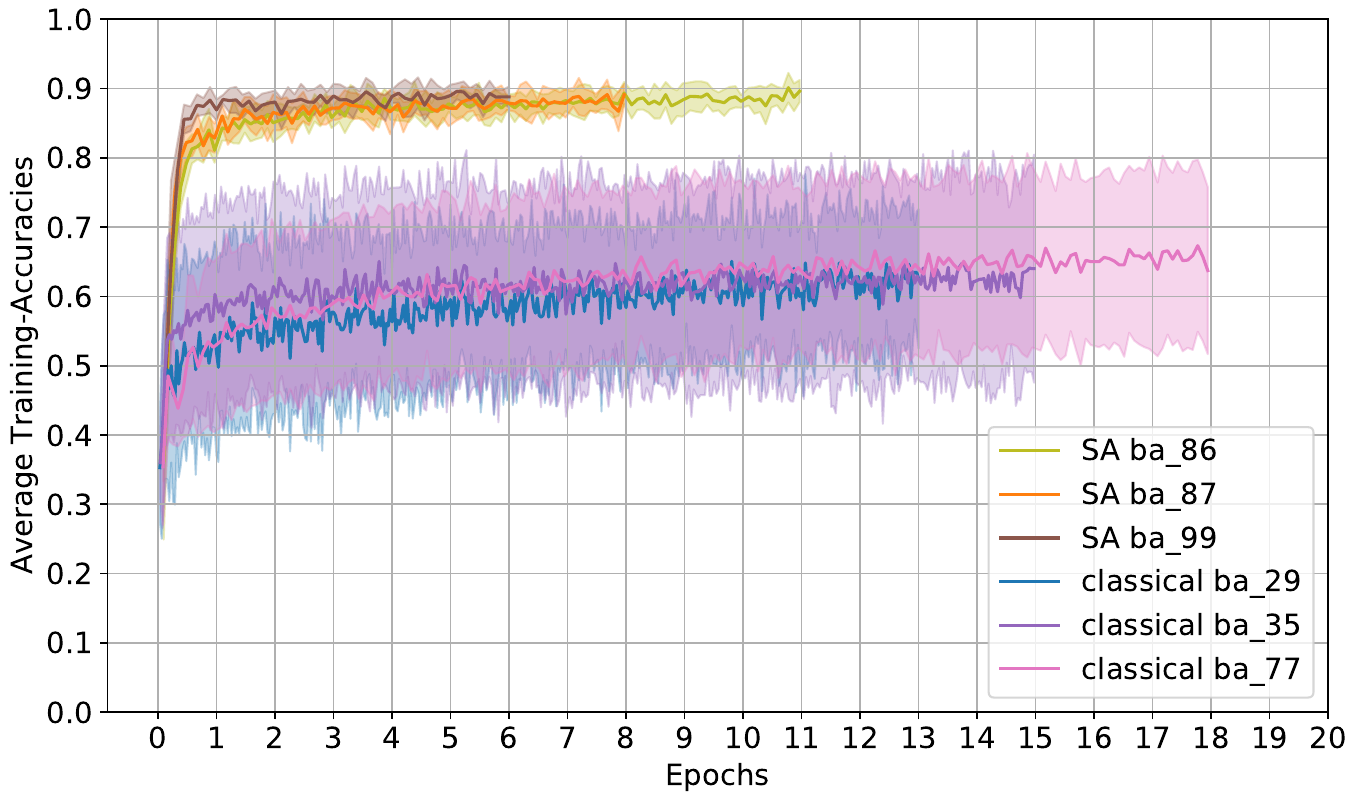}
  \caption{Average training accuracies.}
  \label{subfig:training_acc}
  \end{subfigure}
  \hfill
  \begin{subfigure}{\columnwidth}
  \includegraphics[width=\textwidth]{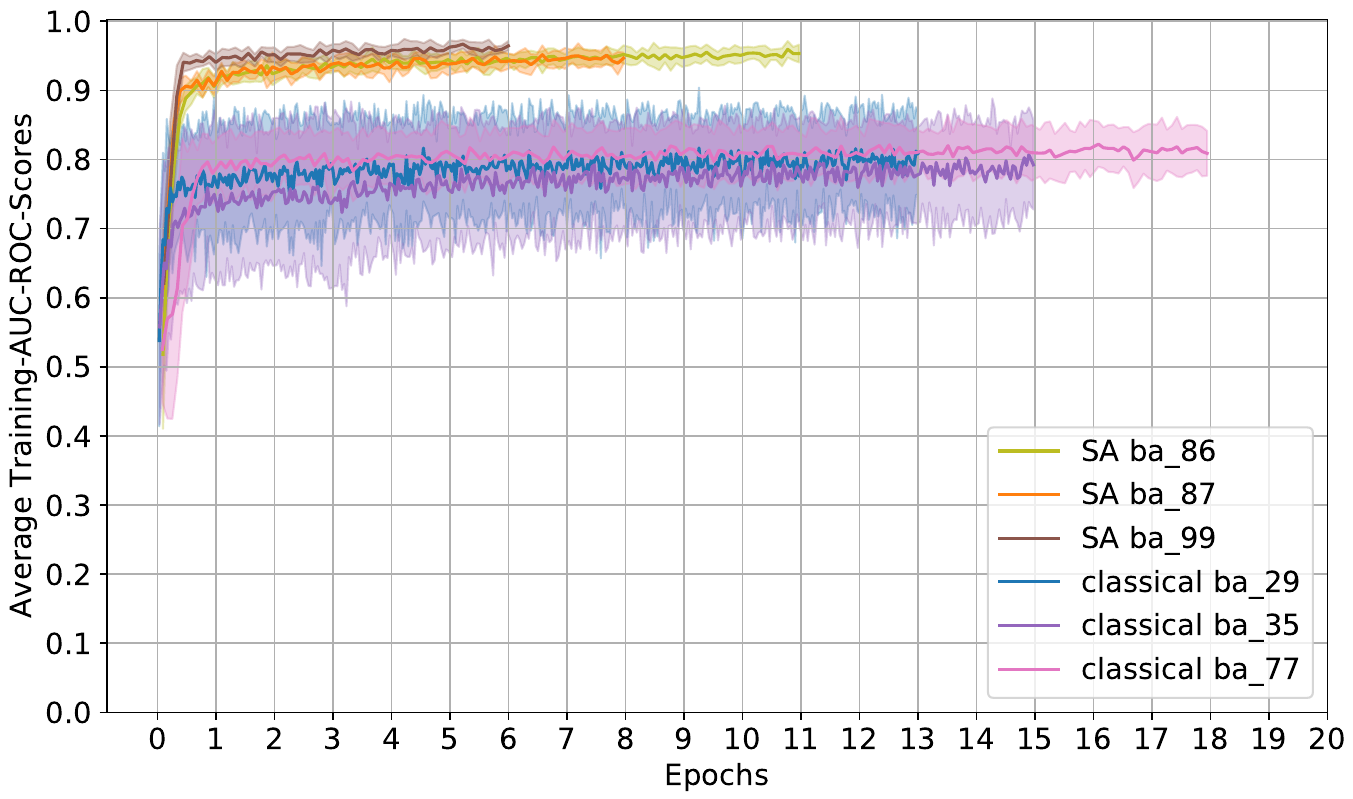}
  \caption{Average training AUC-ROC-Scores.} 
  \label{subfig:training_auc}
  \end{subfigure}
  \caption{Accuracies and AUC-ROC-Scores of SA and classical models with the best identified hyperparameter settings over the course of the training. The solid line shows the average over 10 seeds, the transparent area around it the standard deviation.}
  \label{fig:train}
\end{figure*}

Subsequently, we apply all of the models trained with the respective three best identified hyperparameter configurations to our test set. For each of these models, of which we have 10 per hyperparameter configuration (as we used 10 training seeds), we run the test 10 times, each time using a different test seed. Subsequently, we determine the test values by averaging over these test seeds. The distribution of the test values over the different models can be seen in Fig.~\ref{fig:test}: While the results are generally not optimal, the annealing models on average outperform the classical models and again show more consistent behavior across training seeds.\\

\begin{figure*}[htbp]
  \begin{subfigure}{\columnwidth}
  \includegraphics[width=\textwidth]{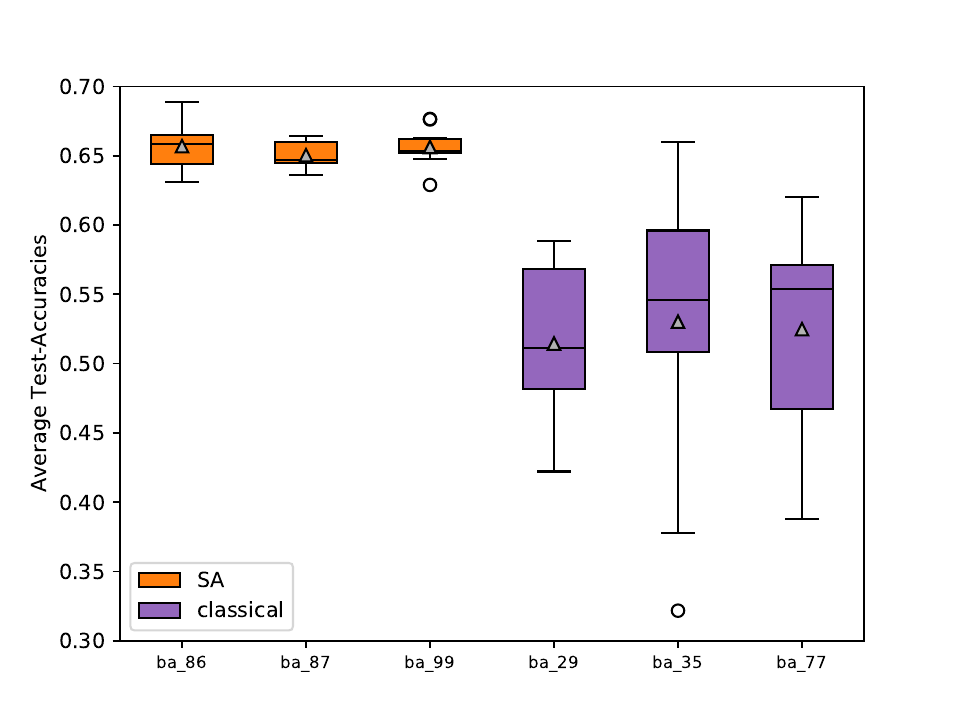}
  \caption{Average test accuracies.}
  \label{subfig:test_acc}
  \end{subfigure}
  \hfill
  \begin{subfigure}{\columnwidth}
  \includegraphics[width=\textwidth]{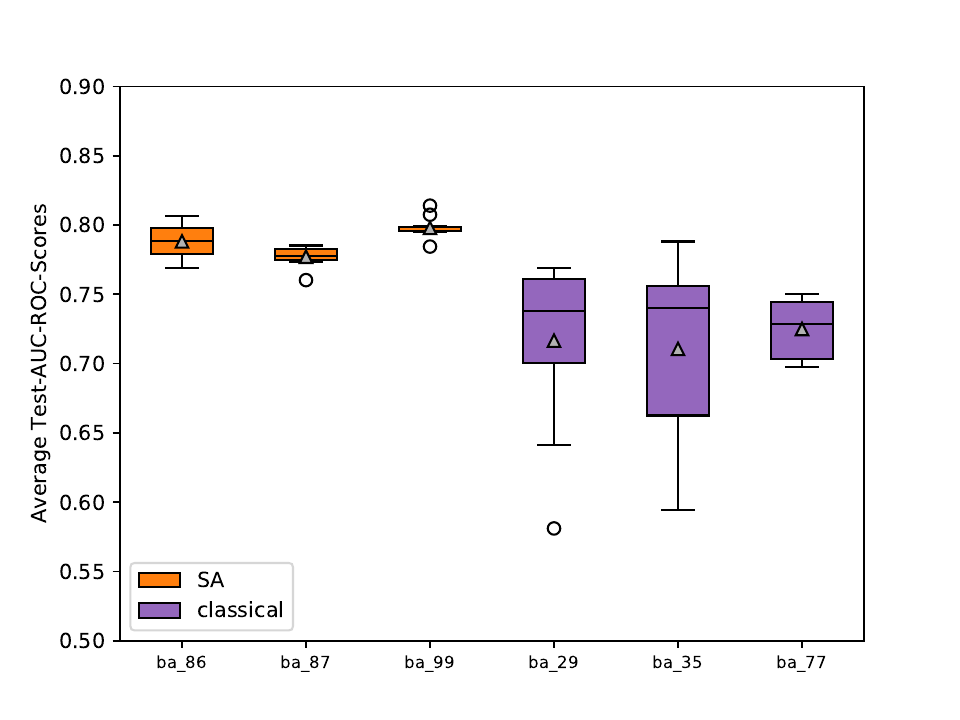}
  \caption{Average test AUC-ROC-Scores.} 
  \label{subfig:test_auc}
  \end{subfigure}
  \caption{Accuracies and AUC-ROC-Scores of SA and classical models with the best identified hyperparameter settings applied to the test set (averaged over 10 test seeds each).}
  \label{fig:test}
\end{figure*}
\section{Discussion and Conclusion}\label{sec:conclusion}
In this work, we have used a hybrid QTL approach including an annealing-based QBM to classify images of a large real-world dataset, namely the COVID-CT-MD dataset. While the classification performance of the approach with under 70\% test accuracy and a test AUC-ROC-Score of only around 80\% is not very high, it on average significantly outperforms a similarly-sized sigmoidal FNN. Also, even though it still takes significantly more wall clock time when using SA, it needs a smaller amount of training epochs to reach this level of performance and is less variable in its performance when using different training seeds. This indicates that, whilst not optimal yet, the approach seems promising for further research into the possible advantages of QML for large-scale image recognition, be it regarding classification performance or execution speed.

Future work should thus include three aspects:

Firstly, possibilities to achieve higher classification performance with the current approach should be investigated. Its suboptimal classification performance is likely due to overfitting on the training dataset, which is neither very large nor diverse, considering it only contains images of a small amount of patients which, when coming from the same person, are very similar. Thus, methods to circumvent this problem, such as employing validation and early stopping, or using another dataset with more different pictures, should be explored.

Secondly, a strong limitation of this work is that so far, we have only used SA to evaluate our approach. Performing experiments on quantum hardware is however a necessary step to determine the actual capability of the approach to enable effectively utilizing near-term available QCs in large-scale image classification. The reason for this is that Quantum Annealers come with a lot of physical properties that might cause their behavior to differ from that of SA: Features like quantum tunneling might have the potential to improve the performance of the approach, while effects of noise, early freeze-outs of the physical dynamics of the annealing process or the “breaking” of entanglement between visible and hidden units might harm it~\cite{venegas2018cross, Benedetti16, Amin18, wiebe2019hidden}. In this context, it might also be interesting to see how our approach compares to the original SEQUENT approach using VQCs and how it compares to a version using supervised classical RBMs instead of the QBM or FNN, to explore the performance of different comparable approaches.

Lastly, while we do suspect that using our pipeline for large-scale image classification is beneficial, due to its employment of feature extraction and further data compression that casts the data into a more information-dense form, we have yet to investigate experimentally how this compares to using a stand-alone QBM on this type of data, without involved pre-processing.

Thus, although the approach presented in this work may seem promising, the question as to whether it will lead to a near-term quantum advantage remains open.\\
 \\

\bibliographystyle{IEEEtran} 
\bibliography{IEEEabrv,references}

\end{document}